\documentclass[conference]{IEEEtran}
\IEEEoverridecommandlockouts
\usepackage{cite}
\usepackage{amsmath,amssymb,amsfonts}
\usepackage{algorithmic}
\usepackage{graphicx}
\usepackage{textcomp}
\usepackage{placeins}
\usepackage[dvipsnames]{xcolor}
\usepackage{float}

\usepackage[numbers,sort&compress]{natbib}
\usepackage{subcaption}
\newcommand{\JC}[1]{{\textcolor{black}{#1}}}

\def\BibTeX{{\rm B\kern-.05em{\sc i\kern-.025em b}\kern-.08em
    T\kern-.1667em\lower.7ex\hbox{E}\kern-.125emX}}
\begin{document}

\title{\textit{Ab initio} Self-consistent GW Calculations in Non-Equilibrium Devices: Auger Recombination and Electron-Electron Scattering\\
}
\author{\IEEEauthorblockN{Leonard Deuschle\IEEEauthorrefmark{1},
Jonathan Backman, Mathieu Luisier, and
Jiang Cao}
\IEEEauthorblockA{Integrated Systems Laboratory,
ETH Zurich,
Switzerland\\
Corresponding Author: \IEEEauthorrefmark{1}dleonard@iis.ee.ethz.ch}}

\maketitle

\begin{abstract}
We present first-principles quantum transport simulations of single-walled carbon nanotubes based on the NEGF method and including carrier-carrier interactions within the self-consistent GW approximation. Motivated by the characteristic enhancement of interaction between charge carriers in one-dimensional systems, we show that the developed framework can predict Auger recombination, hot carrier relaxation, and impact ionization in this type of nanostructures. Using the computed scattering rates, we infer the inverse electron-hole pair lifetimes for different Auger processes in several device configurations.
\end{abstract}


\section{Introduction}
Experimental investigations on semiconducting single-walled carbon nanotubes (SWCNs) have revealed the strong role of electron-electron (e-e) interactions in these devices \cite{PhysRevLett.96.057407, PhysRevB.70.241403, PhysRevLett.101.256804}. First, the measured bandgap is greater than the one predicted by Density Functional Theory (DFT). Second, non-radiative Auger recombination \cite{PhysRevB.70.241403} and avalanche-type processes \cite{PhysRevLett.101.256804} are observed, which are clear signatures that strong e-e interactions may occur. These findings have led to theoretical studies of SWCNs that go beyond one-electron models \cite{Spataru2004}. However, a complete \textit{ab initio} treatment of the e-e interactions in non-equilibrium quantum transport simulations remains missing. Self-consistently treating e-e scattering within the Non-equilibrium Green's function (NEGF) formalism and the GW approximation (scGW) is indeed a formidable challenge. It has only been successfully addressed in small molecules due to the high computational burden associated with such calculations \cite{PhysRevB.77.115333}. In this work, we present a \JC{fully} \textit{ab initio}  \JC{scGW} simulation study of (8,0)-SWCN  devices. We demonstrate that our method naturally accounts for Auger processes under different doping and bias configurations of the SWCN as well as under non-equilibrium conditions. To the best of our knowledge this is the first \textit{ab initio} study of e-e scattering in non-equilibrium devices.

\section{Method}
\begin{figure*}[htbp]
    \centering
        \includegraphics[width=0.8\textwidth]{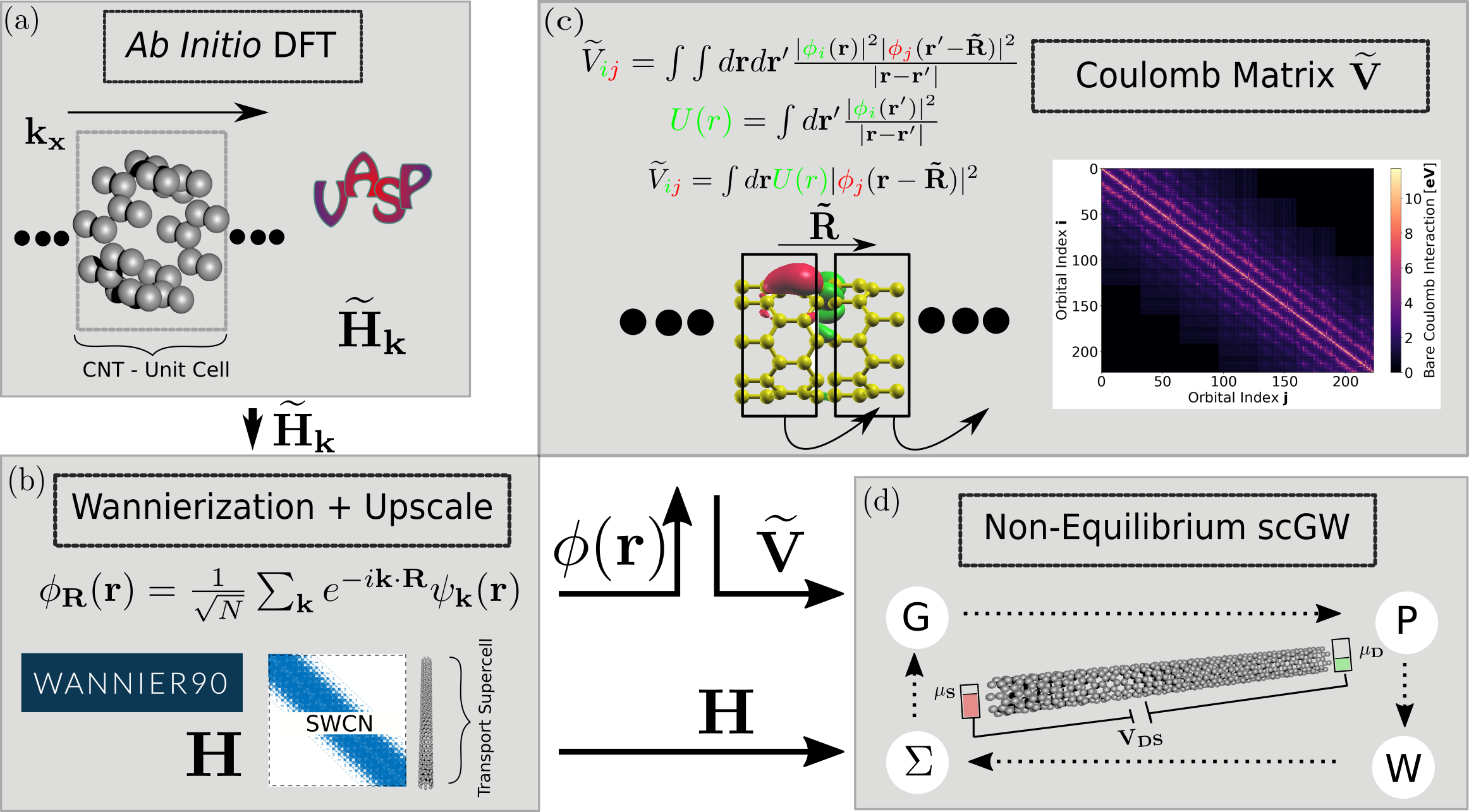}
              \caption{Workflow of the simulation process. (a) Electronic structure calculation with \textit{ab initio} DFT as implemented in VASP [6].  (b) Transformation into a set of maximally localized Wannier functions (MLWFs) with wannier90 [7]. (c) Evaluation of the bare Coulomb potential $\widetilde{V}$ based on the real-space Wannier functions. (d) Calculation of the Non-Equilibrium scGW-loop. After convergence device observables can be extracted. Here, the calculation is performed for a (8-0)-SWCN with a diameter of 6.29 $ \mathrm{\AA}$ and containing 896 atoms. The nanotube is about 12 nm long.}
            \label{fig:flowchart}
\end{figure*}
\subsection{Hamiltonian Generation}
We first perform a DFT calculation of the (8,0)-SWCN unit cell with VASP \cite{PhysRevB.47.558} using GGA-PBE pseudo-potentials,
assuming that the nanotube transport direction is periodic. We then use wannier90 \cite{Mostofi2014} to transform the plane-wave basis into a set of maximally localized Wannier functions (MLWF) and upscale the Hamiltonian to the size of the device of interest. We created a SWCN with a diameter of 6.29 $\mathrm{\AA}$, a length of roughly 12 nm, and made of 896 atoms with one single orbital per atom. (Fig.~\ref{fig:flowchart}a-b).

\subsection{Non-Equilibrium scGW}
The following equation must be solved to obtain the retarded Green's function $G^R$
\begin{equation}
\left(E-H-\Sigma^{R}_{B}(E)- \Sigma^{R}_{GW}(E)\right) \cdot G^{R}(E)  = I,
\label{eq:GR}
\end{equation}
where $E$ is the energy vector, $H$ is the Hamiltonian matrix and $\Sigma^{R}_{B}$ and $\Sigma^{R}_{GW}$ are the retarded boundary and GW self-energies, respectively. $I$ denotes the identity matrix.
The lesser and greater Green's function $G^{\lessgtr}$ are given by 
\begin{equation}
\label{eq:lessgtrGLGtilda}
   G^{\lessgtr}(E) = G^R(E)\cdot \Sigma^{\lessgtr}(E) \cdot {G^{R}}^{\dagger}(E).
\end{equation}
Here, $\Sigma^{\lessgtr}$ denotes the sum of the lesser and greater boundary and GW self-energies.
The retarded, lesser, and greater GW self-energies are given by
\begin{equation}
\label{eq:sigr}
\Sigma^{\lessgtr}_{GW,ij} = i \int dE' \lbrack G^{\lessgtr}_{ij}(E')W^{\lessgtr}_{ij}(E-E') \rbrack,
\end{equation}

\begin{align}
    \Sigma_{GW,ij}^{R}= i\int  dE' &\lbrack G^{R}_{ij}(E')W^{<}_{ij}(E-E') \nonumber
    \\
    &+ G^{<}_{ij}(E')W^{R}_{ij}(E-E') \nonumber
    \\
    \label{eq:siglg}
    &+ G^{R}_{ij}(E')W^{R}_{ij}(E-E')\rbrack,
\end{align}
where $W$ is the screened interaction. The self-energy matrix elements between orbital index $i$ of an atom situated at position $\mathrm{\mathbf{R}_i}$  and another one $j$ at $\mathrm{{\mathbf{R}_j}}$ in Eqs.~\eqref{eq:sigr} and \eqref{eq:siglg} are found by performing the convolution in an element-wise fashion.
Note that all energy convolutions are explicitly calculated in the energy space, not through Fourier transform.
The retarded, lesser, and greater screened interactions are calculated using
\begin{equation}
\label{eq:wr}
W^R (E)= \lbrack I - \widetilde{V}P^R(E) \rbrack^{-1} \widetilde{V},
\end{equation}
\begin{equation}
W^{\lessgtr}(E) = W^R(E)P^{\lessgtr}(E){W^{R}}^{\dagger}(E).
\end{equation}

The equations for the components of the irreducible polarization $P$ in the energy domain read
\begin{equation}
P^{\lessgtr}_{ij} = - i \int dE' \lbrack G^{\lessgtr}_{ij}(E')G^{\gtrless}_{ji}(E'-E) \rbrack,
\end{equation}
\begin{align}
    P_{ij}^{R} = -i \int dE' &\lbrack G^{R}_{ij}(E')G^{<}_{ji}(E'-E) \nonumber
    \\
    &+ G^{<}_{ij}(E')G^{A}_{ji}(E'-E)\rbrack.
\end{align}

To compute Eq.~\eqref{eq:wr} the bare Coulomb matrix elements are needed.
The real-space representation of the MLWF is used to calculate the Coulomb matrix elements $\widetilde{V}'_{ij}$: 
\begin{equation}\label{eq:Coulomb_matrix}
    \widetilde{V}'_{ij} = \int \int d\mathbf{r}d\mathbf{r'} \frac{|\phi_i(\mathbf{r})|^2|\phi_j(\mathbf{r'})|^2}{|\mathbf{r} - \mathbf{r'}|}.
\end{equation}
In Eq.~\eqref{eq:Coulomb_matrix}, the $\phi_i$'s denote a single MLWF in a real-space basis. 
The approach outlined in \cite{PhysRevB.77.115333} is used to first compute an electrostatic potential induced by one of the two wavefunctions and then integrating the potential weighted by the second wavefunction over space. The matrix must be upscaled to the device structure. Since the wavefunctions obey lattice periodicity, off-diagonal blocks can be computed by shifting one wavefunction into a different cell and then computing the resulting integral (Fig.~\ref{fig:flowchart}c). 
To correct for the truncation scheme used in the scGW, the Coulomb matrix is embedded in a dielectric environment $\widetilde{V}$=$\widetilde{V}' / \epsilon$.
The $\epsilon$ is chosen such that the equilibrium nanotube bandgap matches the one obtained from a VASP  DFT + $\mathrm{G_0W_0}$ calculation.

\JC{To fulfill the conservation laws,} the scGW is performed within the self-consistent Born approximation (SCBA) (Fig.~\ref{fig:flowchart}d). Only the diagonal elements of the polarization and of the GW self-energy are kept \JC{in this study} to minimize the computational burden. \JC{However, we have verified that the inclusion of off-diagonal elements does not qualitatively alter our results.} Once the convergence \JC{of SCBA} is reached, several device observables (i.e., local density-of-states (LDOS), charge densities, and spectral current) can be computed. Additionally, it was confirmed that current conservation is satisfied for the converged solution along the transport direction of the structure.
\subsection{e-e Scattering Rates}
In the NEGF formalism, the energy-resolved in- ($\mathcal{R}_{e-e}^{in}$) and out-scattering ($\mathcal{R}_{e-e}^{out}$) rates are computed with

\begin{align}
\mathcal{R}_{e-e}^{in} (E) &= \frac{1}{2\pi \hbar}\mathrm{tr}\{\Sigma^{<}(E)G^{>}(E)\}
\label{eq:Scat-in}
\\
\mathcal{R}_{e-e}^{out} (E) &= \frac{1}{2\pi \hbar}\mathrm{tr}\{\Sigma^{>}(E)G^{<}(E)\}.
\label{eq:Scat-out}
\end{align}

To infer the inverse annihilation lifetimes of excited electron-hole pairs in an investigated Auger process, all quantites in Eqs.~\eqref{eq:Scat-in}-\eqref{eq:Scat-out} are integrated over the relevant energy window and over the spatial dimension over which the process occurs. They are then normalized by the number of available electron-hole pairs in the considered energy range.

\section{Results}
\begin{figure}[ht]
\centerline{\includegraphics[width = 0.32\textwidth]{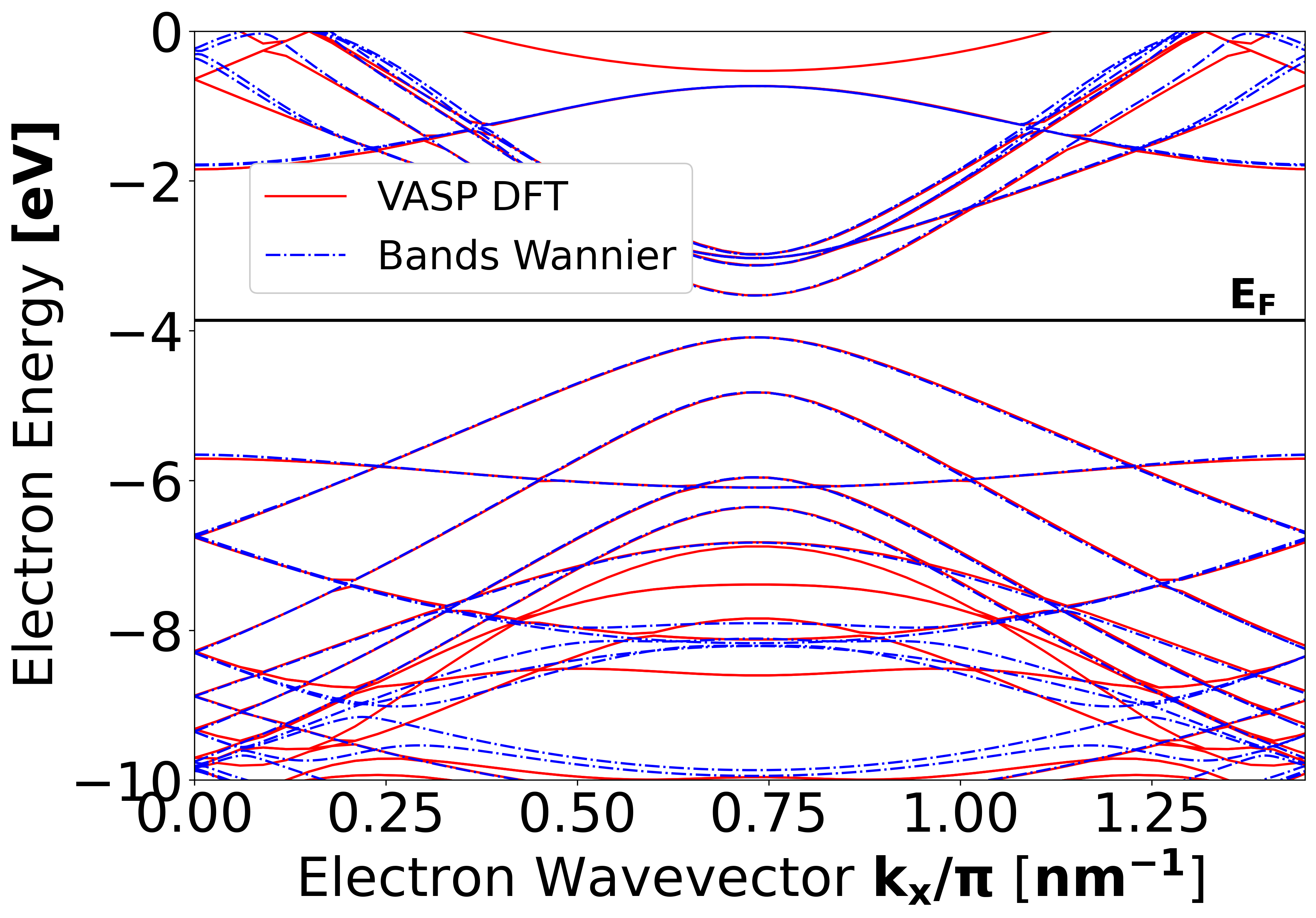}}
\caption{DFT- (solid) and MLWF (dashed) -based bandstructure of the (8-0)-SWCN unit cell.}
\label{fig:VAPS_V_Wannier}
\end{figure}

\begin{figure}[htbp]
     \centering
        \begin{subfigure}[t]{0.33\textwidth}
         \centering
         \includegraphics[height = 4.25 cm, width = \textwidth]{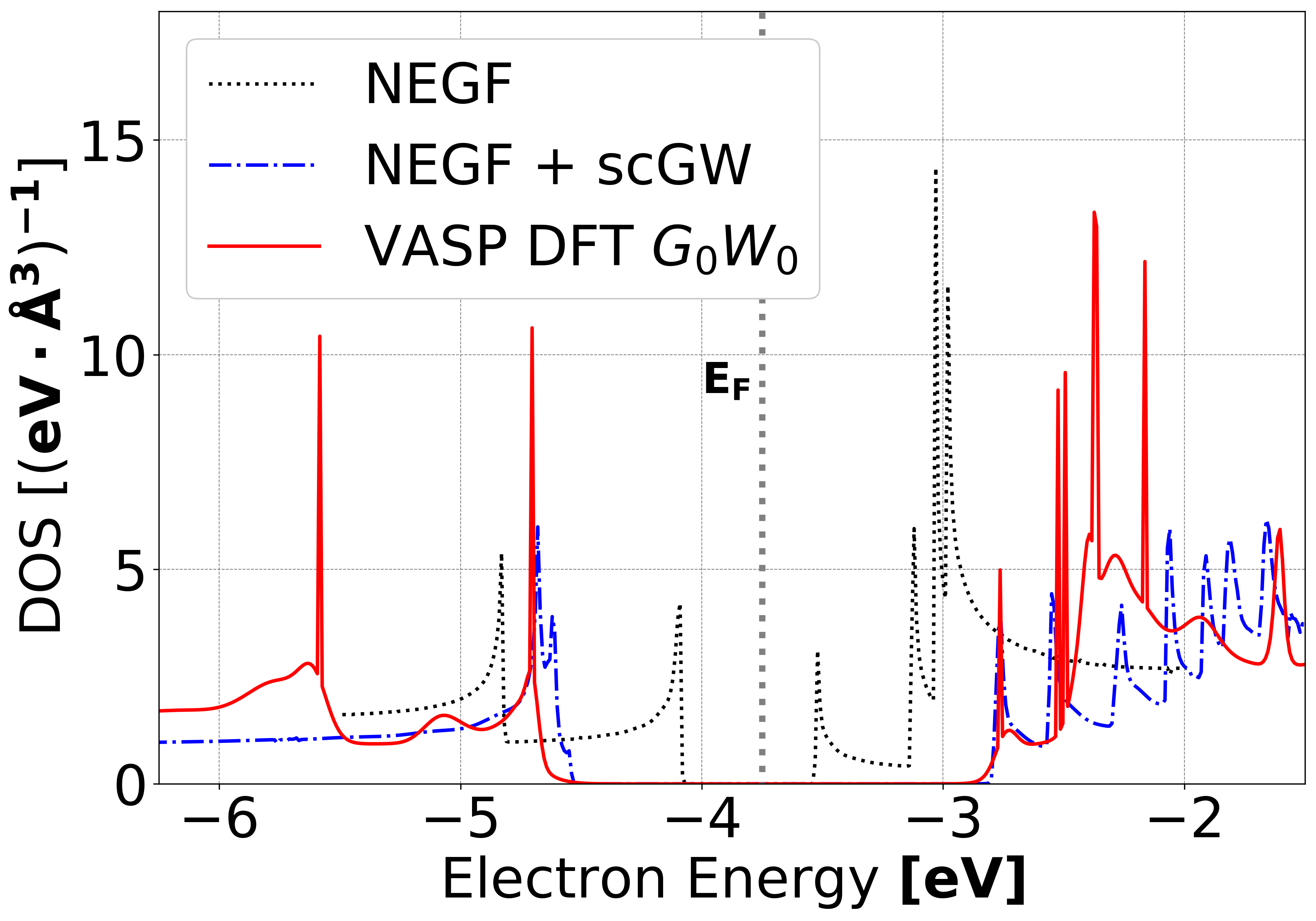}
         \caption{}
         \label{fig:LDOS_line}
     \end{subfigure}%
     \hspace{0em}%
     \begin{subfigure}[t]{0.33\textwidth}
        \centering
         \includegraphics[height = 4.25 cm, width = \textwidth]{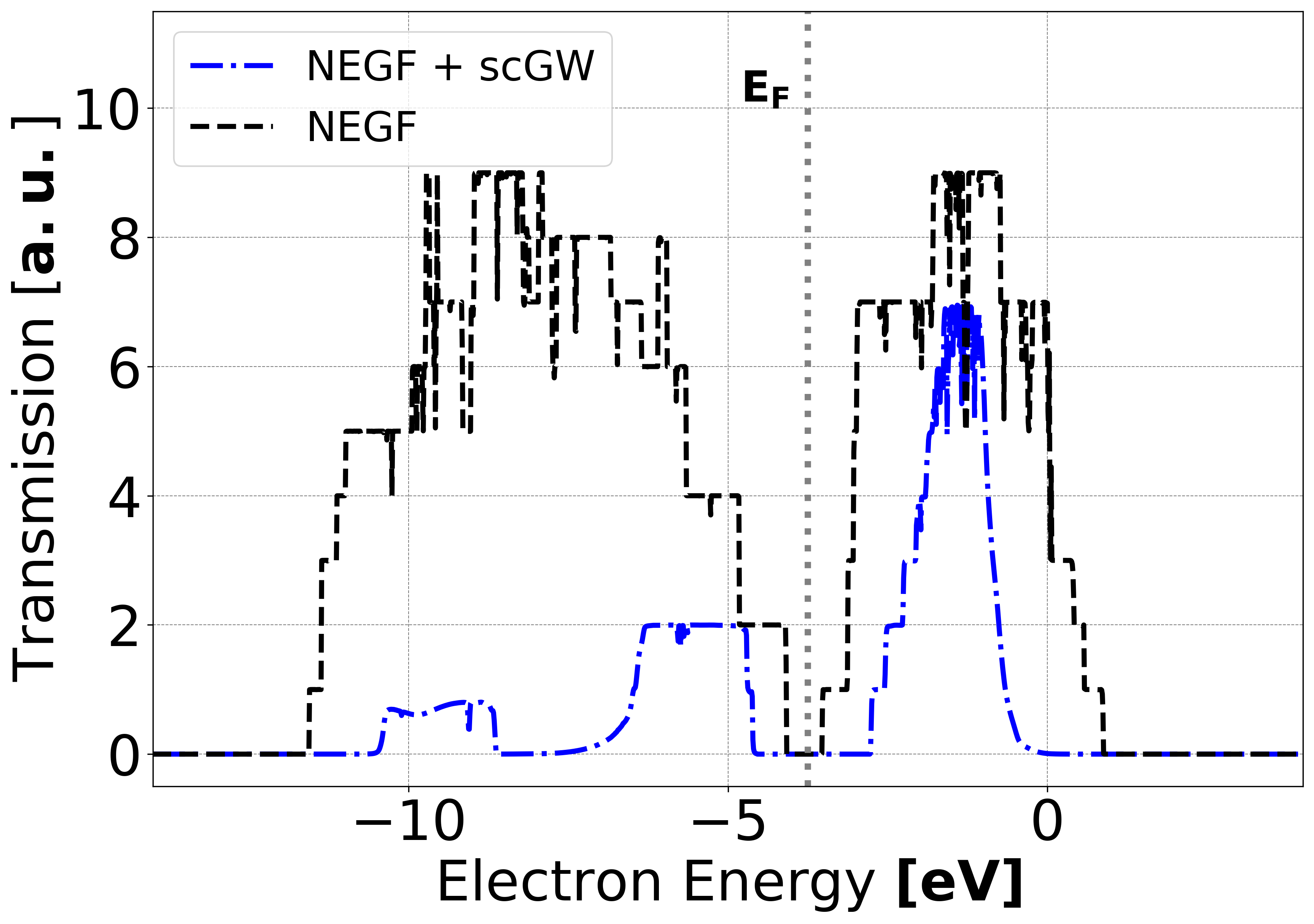}
         \caption{}
         \label{fig:LDOS_BALL}
     \end{subfigure}%
     \hspace{0em}%
     \begin{subfigure}[t]{0.33\textwidth}
         \centering
         \includegraphics[height = 4.25 cm, width = \textwidth]{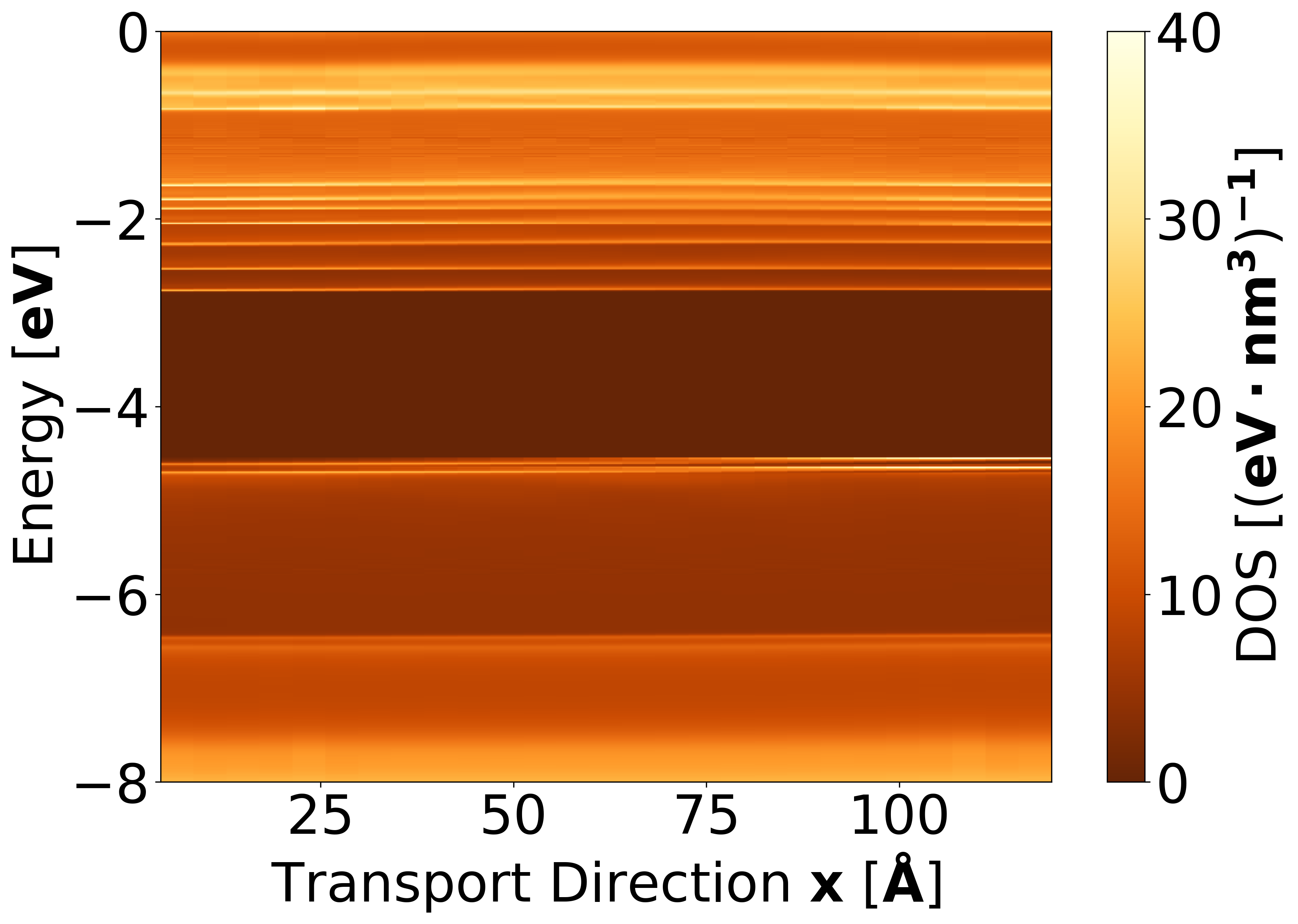}
         \caption{}
         \label{fig:LDOS_GW}
     \end{subfigure}%
     \caption{(a) Density-of-states comparison between NEGF (black dotted line), NEGF + scGW (blue dash-dots), and the VASP $G_0W_0$ calculation (red solid line). (b) Transmission function for the NEGF (black dotted line) and NEGF + scGW (blue dash-dots) case. (c) Position-dependent DOS through the CNT with a flat potential under equilibrium conditions with sc-GW.}
     \label{fig:mid-row}
\end{figure}

\begin{figure}[htbp]
     \centering
     \begin{subfigure}[t]{0.33\textwidth}
         \centering
         \includegraphics[height = 5.15 cm, width = \textwidth]{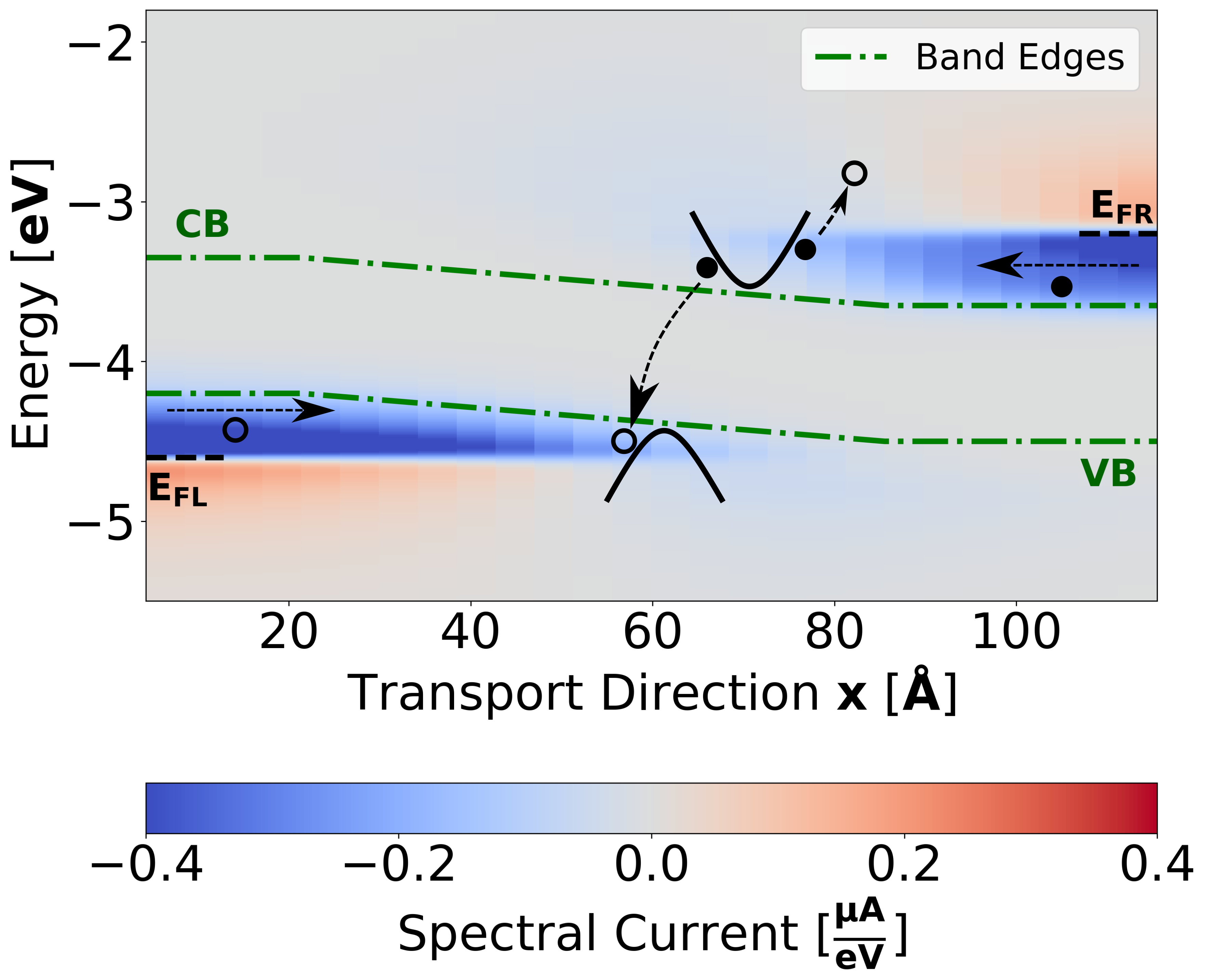}
         \caption{}
         \label{fig:PN}
     \end{subfigure}%
     \hspace{0em}%
     \begin{subfigure}[t]{0.33\textwidth}
         \centering
         \includegraphics[height = 5.15 cm, width = \textwidth]{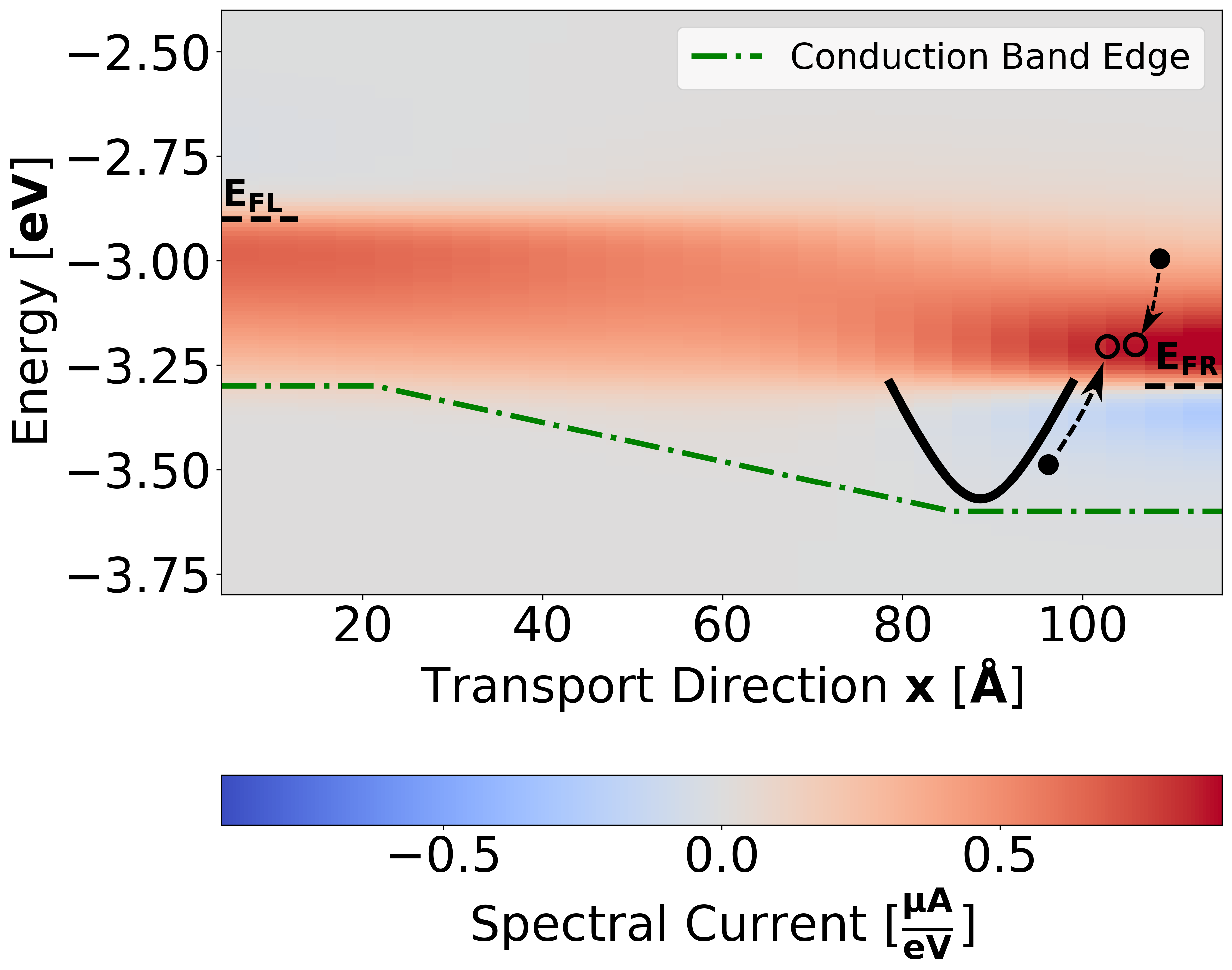}
         \caption{}
         \label{fig:NN}
     \end{subfigure}%
     \hspace{0em}%
     \begin{subfigure}[t]{0.33\textwidth}
         \centering
         \includegraphics[height = 5.15 cm, width = \textwidth]{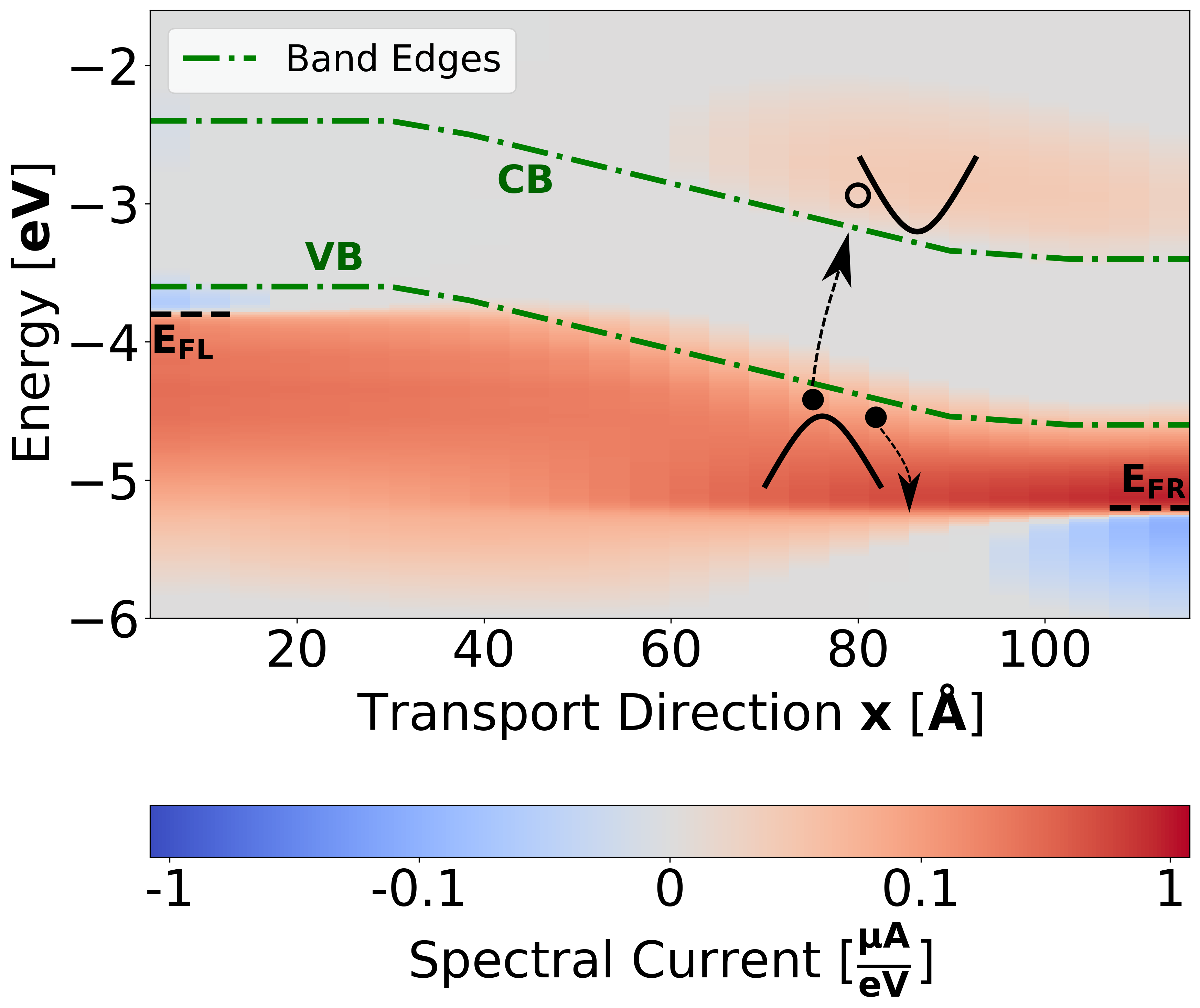}
         \caption{}
         \label{fig:PP}
     \end{subfigure}
        \caption{Spectral current of the SWCN at different source/drain doping concentrations under non-equilibrium condition. The Fermi levels are indicated with dotted lines in the left and right contacts. The green dash-dots indicate the approximate conduction- and valence-band edges, respectively. The drawings in solid black lines illustrate the energy-conserving e-e scattering processes. (a) PN setup with Auger recombination. A voltage $\mathrm{V = 1.4 V}$ is applied, assuming a linear potential drop. (b)  NN case with hot carrier relaxation at $\mathrm{V = 0.4 V}$. (c) PP configuration exhibiting impact ionization at $\mathrm{V = 1.2V}$.}
        \label{fig:results}
\end{figure}
\begin{figure}[ht]
\centerline{\includegraphics[width = 0.35\textwidth]{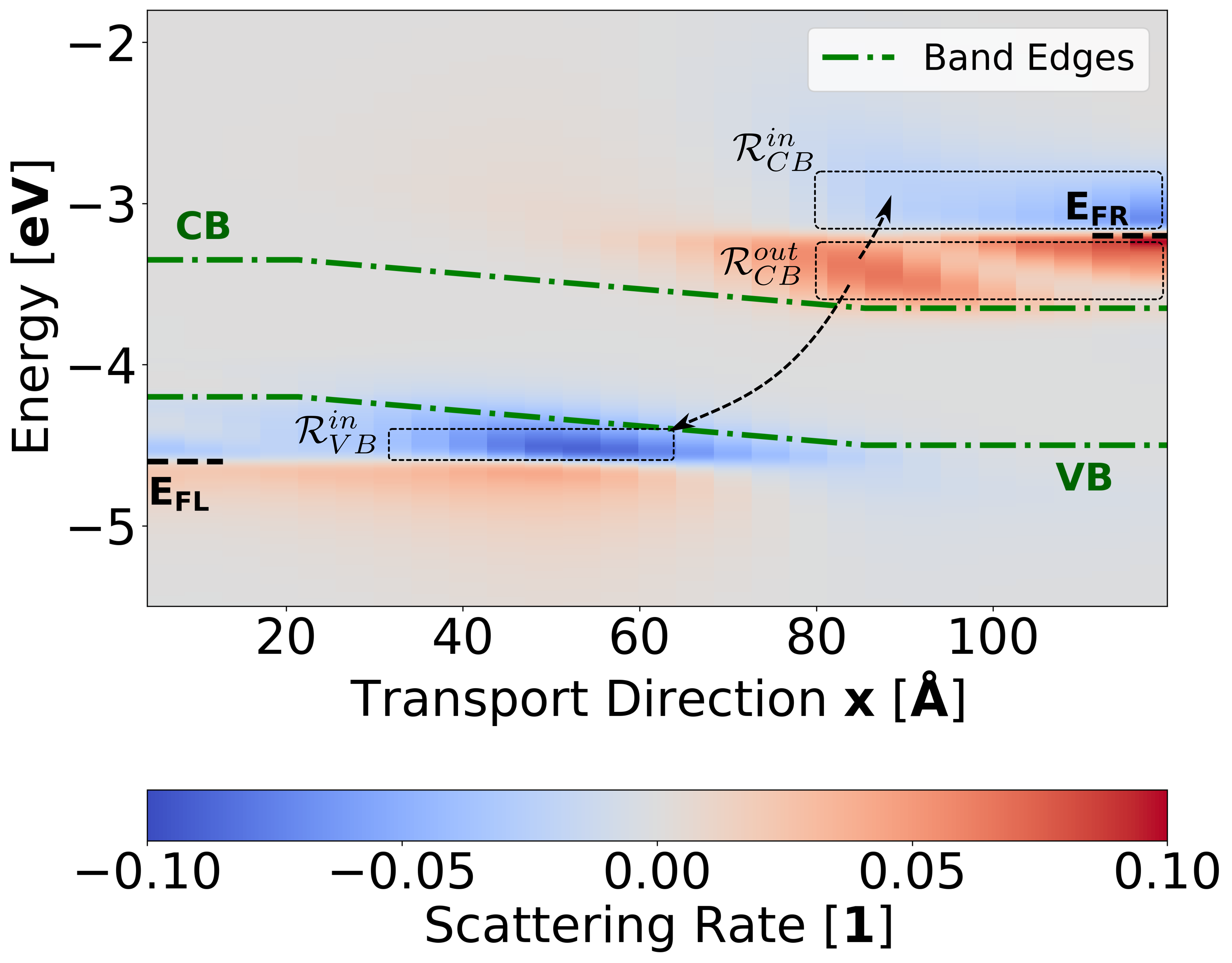}}
\caption{Computed energy-resolved scattering rates in the PN configuration. Near the right contact, conduction band electrons are scattered to higher eneregies and recombine with empty states in the valence band. }
\label{fig:PN_scat}
\end{figure}
A bandstructure comparison between DFT and MLWF is reported in Fig.~\ref{fig:VAPS_V_Wannier}. Excellent agreement is observed for the valence bands as well as the first 4 conduction bands, which were set as targets. Using $\epsilon$=1.1, the bandgap as extracted from our scGW LDOS at equilibrium condition matches the one obtained with VASP $\mathrm{G_0W_0}$ at 1.92 $\mathrm{eV}$, as shown in Fig.~\ref{fig:LDOS_line}. Figures~\ref{fig:mid-row}b-c indicate that the increase in bandgap \JC{after scGW} is accompanied by a change in the shape of the transmission function \JC{due to the GW self-energy}. We then apply an external potential and drive the SWCN out-of-equilibrium. The Fermi levels are adjusted relative to the valence (VB) and conduction band (CB) edges to create PN-, NN- and PP-like structures. Figure~\ref{fig:results} reports the position- and energy-resolved current density for the SWCN at different biases and contact doping profiles. In the PN-like structure (Fig.~\ref{fig:PN}), the electrical current is injected into the CB from the right contact. In the middle of the device carriers recombine with the available hole states in the valence band (Auger recombination), that come from the left contact. In the NN case (Fig.~\ref{fig:NN}), high-energy electrons collide with low-energy electrons on the right side, exchanging energy during this process and giving rise to thermalization effects. Finally, in the PP case, depicted in Fig.~\ref{fig:PP}, impact ionization can be observed due to the large bias applied. Electrons in the valence band release kinetic energy and excite other electrons  to empty states in the conduction band.

The in- and out-scattering rates observed in the PN-like structure are further investigated in Figure~\ref{fig:PN_scat}. The red shaded area in the conduction band depicts all out-scattering electrons in the energy range. The blue shade above represents the in-scattered electrons inside the conduction band. Subtracting the blue area from the red yields the total number of transitions from conduction to valence band (Auger recombination). This number is then normalized by the total number of available electron-hole pairs.
The calculated Auger recombination (AR), impact ionization (II), and inverse electron-hole pair lifetimes from our simulations are summarized in Tab.~\ref{Tab:scatrates}.
\begin{table}[ht]
\small
\centering
   \begin{tabular}{|l|l|c|}
   \cline{1-3}
   Doping & Process & Inv. Lifetime [$\mathrm{ps}^{-1}$]  \\ \cline{1-3}
   \hline
   PN & AR & 1.41 \\ \cline{1-3}
   PP & II & 0.54  \\ \cline{1-3}
   NN & e-e & 0.40 \\ \cline{1-3}
   \end{tabular}
   \caption{Calculated scattering rates for the 3 different configurations.}
         \label{Tab:scatrates}
\end{table}
The value in the AR case shows good agreement with time-resolved fluorescence measurements \cite{PhysRevB.70.241403}.

\section{Conclusions}
We have shown that our recently developed  \textit{ab initio} scGW method can accurately model e-e interactions in SWCNs driven out-of-equilibrium.  A very wide range of scenarios can be investigated, from Auger recombination to hot carrier relaxation and impact ionization. This method can be readily applied to any (quasi-)1D system, e.g., nanowires and nanoribbons. As the next step, off-diagonal elements will be added and the numerics will be improved to treat larger device structures.

\section*{Acknowledgment}

This work was supported by the Swiss National Science Foundation (SNSF) under grant $\mathrm{n^\circ}$ 209358 (QuaTrEx). We acknowledge support from CSCS under project s1119. J.C. acknowledges funding from the European Union under the Marie Skłodowska-Curie grant No. 885893.

\FloatBarrier
\bibliographystyle{IEEEtran}
\bibliography{bibliography}


\end{document}